\documentstyle[aps,prl,multicol,epsf]{revtex}
\begin{document}
\title{\bf Self-Organized Criticality and Universality in a Nonconservative 
Earthquake Model}

\author{Stefano Lise and Maya  Paczuski}
\address{Department of Mathematics, Huxley Building, Imperial College
of Science, Technology, and Medicine, London UK SW7 2BZ \\}
\date{\today}

\maketitle %\parskip 2ex

\begin{abstract}
 
 We make an extensive numerical study of a two dimensional
 nonconservative model proposed by Olami-Feder-Christensen to describe
 earthquake behavior.  By analyzing the distribution of earthquake
 sizes using a multiscaling method, we find evidence that the model is
 critical, with no characteristic length scale other than the system
 size, in agreement with previous results.  However, in contrast to previous
 claims, we find convergence to universal behaviour
 as the system size increases, over a range of values of the dissipation
 parameter, $\alpha$.  We also find that both ``free'' and ``open''
 boundary conditions tend to the same result.
 Our analysis indicates that, as $L$ increases, the behaviour
 slowly converges toward a power law distribution of earthquake sizes
 $P(s) \sim s^{-\tau}$ with exponent $\tau \simeq 1.8$. The universal
 value of $\tau$ we find numerically agrees quantitatively with the empirical
 value ($\tau=B+1$) associated with the Gutenberg-Richter law.

\end{abstract}


%%%%%%%%%%%%%%%%%%%%%%%%%%%%%%%%%%%%%%%%%%%%%%%%%%%%%%%%%%%%%%%%
\begin{multicols}{2}

\section{Introduction}

Earthquakes may be the most dramatic example of self-organized
criticality (SOC) \cite{soc,review} that can be observed by humans on
earth.  Most of the time the crust of the earth is at rest, or
quiescent. These periods of stasis are punctuated by sudden, thus far
unpredictable bursts, or earthquakes.  According to the empirical
Gutenberg-Richter (GR) law \cite{gutenberg}, the distribution of
earthquake events is scale-free over many orders of magnitude in
energy.  The GR scaling extends from the smallest measurable
earthquakes, which are equivalent to a truck passing by, to the most
disastrous that have been recorded where hundreds of thousands of
people have perished.

The relevance of SOC to earthquakes was first pointed out by Bak and
Tang \cite{bak-tang}, Sornette and Sornette \cite{sornette}, as well
as Ito and Matsusaki \cite{ito}.  According to this theory, plate tectonics
provides energy input at a slow time scale into a spatially extended,
dissipative system that can exhibit breakdown events via a chain
reaction process of propagating instabilities in space and time.  The
GR law arises from the system of driven plates building up to
a critical state with avalanches of all sizes.  These above-mentioned
authors used a spatially
extended (but conservative) cellular automata model as a prototype
resembling earthquake dynamics which gave a power law distribution of
avalanches, or earthquakes.  This was followed by a study using
block-spring models \cite{burridge} by Carlson and Langer
\cite{carlson-langer}, who found characteristic earthquake sizes,
rather than asymptotic critical behavior.  Later studies of a continuous
``train'' block-spring model by de Sousa Vieira \cite{vieira}
recovered criticality. The train model describes a driven elastic
chain sliding over a surface with friction.  It has been conjectured
to be in the same universality class as interface depinning and a
model of avalanches in granular piles, which agrees with numerical
simulation results \cite{universal}.

 Several groups made lattice representations of the block-spring
model.  These models were nonconservative \cite{takayasu,nakanishi}
and were driven uniformly, but did not display SOC.  Then Olami,
Feder, and Christensen (OFC) introduced a nonconservative model on a
lattice that displayed SOC \cite{ofc}.  In this simplified earthquake
model, sites on a lattice are continuously loaded with a force.  After
a threshold force is reached, the sites transfer part of their force
to their local neighborhood when discharging.  Each discharge event is
accompanied by a local loss in accumulated force from the system, when
the force on each element is reset to zero.  A uniform driving force
is slowly applied to all the elements and the model is completely
deterministic.  This conceptually simple and seemingly numerically
tractable model reproduces some of the qualitative phenomenology of
the statistics of earthquake events such as power law behavior over a
range of sizes, intermittency or clustering of large events
\cite{ofc}, and lack of predictability \cite{pepke}.

Although SOC and this type of modelling approach has been more or less
accepted as a reasonable description of the phenomena of earthquakes
(see for example Ref. ~\cite{turcotte} and references therein), the
OFC model itself has had a controversial existence, both on the
theoretical \cite{klein,lise,french,grass1} and numerical side
\cite{ofc,kertez,socolar,grass2,middleton,corral}.  The initial
numerical studies found that the distribution of earthquake sizes
obeyed finite size scaling (FSS) over the range of system sizes that
could be studied at the time \cite{ofc}.  This placed the
nonconservative model into the framework of standard critical
behavior.  However, these simulations also indicated that there was no
universality.  In particular the exponents characterizing the power
law distributions appeared to vary with both the dissipation
parameter, $\alpha$, and the form of boundary condition.  If this were
the case then the OFC model would be very different from familiar
critical systems where most microscopic details are irrelevant and
have no effect on critical coefficients.  In fact an argument was made
\cite{ofc} that one should not expect universal behavior in far from
equilibrium critical phenomena.  If this were correct, it would
drastically limit the application of any known theoretical tools to
these problems.

Another strange aspect was that the dimension $D$ characterizing the
scaling of the cutoff in the earthquake size distribution was found
numerically to be larger than two. This is inconsistent with the fact
that each site can only discharge a finite number of times in an
earthquake event, requiring $D \leq 2$ for the two dimensional model
\cite{klein}.  This last result together with the strange lack of
universality cast some doubt on whether the OFC model was actually
critical or just close to being critical, with some large as yet
undetermined length scale beyond which the earthquake distribution
would always be cut off.   Hwa and Kardar as well as
Grinstein and collaborators postulated that conservation of the
quantity being transported was required for criticality
\cite{hwa-kardar}, but the theoretical arguments made do not take into
account SOC phenomena such as avalanches and long-term memory
associated with the self-organization process (for more details see
\cite{paczuski}).  The fact that the random neighbor version of the
nonconservative OFC model is never critical but has an essential
singularity as the conservative limit is approached
\cite{french,grass1} has added to the mystery surrounding the behavior
of the lattice model.

In a previous large scale numerical simulation study of the model
discussed here, Grassberger \cite{grass2} also claimed that the model
was critical but found that some of the conclusions of OFC ``have to
be modified considerably.''  He argued that the probability
distribution of earthquake sizes does not show ordinary FSS over the
larger range of systems he was able to study, and ``conjecture(d) that
the cutoff of $P(s)$ becomes a step function for $L\rightarrow
\infty$,'' although he did not present direct numerical evidence of this.

There are a number of important, unresolved questions about the
behavior of the model, which have enormous implication for any type of
eventual theoretical understanding.  Is the nonconservative model on a
lattice (or for fixed connectivity matrix)
critical?  If so, is the critical behavior of the model
universal over a range of values of $\alpha$, or for different
boundary conditions?  Is it described by a
power-law distribution at all?  Are there any other universal quantities?
What type of data analysis
technique besides FSS would be useful to extract the large scale
behavior of the nonconservative model?  Our numerical study and
analysis will address those issues and answer those questions.

\subsection{Summary}
In the first section we review the definition of the OFC model and
present some numerical data for the distribution of earthquake sizes
using standard FSS.  For the range of lattice sizes we have simulated
(up to linear size $L=512$), our data confirm the lack of apparent FSS
in the model, particularly in the cutoff region.  In Section III, we
present an extensive set of results using a multiscaling method.
We analyze
the rescaled probability distribution, $\frac{\log P(s)}{\log L}$
\cite{note}, in
terms of the quantity $D_{av}\equiv \log s/\log L$, with $s$ being the
size of an earthquake.  We observe that there are no avalanches with
$D_{av}$ larger than two, consistent with the bound imposed on the
cutoff $s_{co}$ (see previous discussion).  By analyzing how this
distribution behaves for different values of the nonconservation
parameter, $\alpha$, and system size, $L$, we show how the multiscaled
probability distribution tends to converge to a universal curve as $L$
increases.  The direction of convergence on increasing $L$ changes  as
$\alpha$ varies enabling us to put fairly firm limits on the
asymptotic curve.  The model appears not be to described at
all by FSS.  However,  for $s<s_{co}$ the distribution converges
toward a power law with a universal exponent $\tau \simeq 1.8$ over a
range of $\alpha$ values.  Moreover the cutoff in this distribution
becomes very sharp as $L$ increases and its behavior indicates that
$s_{co} \rightarrow {\rm const}(\alpha) L^2$ as $L\rightarrow \infty$.
In Section IV we summarize our main conclusions.

\section{Definition of the Model}
We consider a two-dimensional square lattice of $L \times L $ sites.
At each site $i$, a force $F_i$ is assigned to be a real variable.
Initially, the force at each site is chosen randomly from the uniform
distribution between 0 and 1. The dynamics proceeds by two steps in
the limit of infinite time scale separation between the slow drive,
representing motion of the tectonic plates, and the earthquake process
\cite{ofc}.

\begin{enumerate}
\item{Increase  the force at all sites: Find the largest force $F_{max}$ 
  in the system and increase  the force at all sites by the same 
  amount $1- F_{max}$.}

\item{Relax all unstable sites, i.e. sites with $F_i \geq 1$: The force
  of an unstable site is reset to zero,  $F_i \rightarrow 0$, and a  
  fraction of it, $\alpha F_i$, is distributed  to each of its four 
  nearest neighbours, $F_{nn} \rightarrow F_{nn} + \alpha F_i$.   
  This step is repeated in a parallel update until there are no 
  unstable sites left.}

\end{enumerate}
  This two step rule is iterated indefinitely.  The sequence of
  toppling events (step 2) between application of the uniform drive (step 1)
  defines an avalanche or earthquake.  Since only a fraction,
  $4\alpha$, of the force is redistributed in each toppling, the model
  is nonconservative for $\alpha < 1/4$.

To completely define the model we need to specify the boundary
conditions, by defining the dissipation parameter,
$\alpha$, at the sites on the boundaries ($\alpha _{bc}$).
As in OFC, we consider both ``free'' and ``open'' boundary conditions.  
The sites on the boundaries of the system can be considered to be bounded by 
fictitious sites with $F_i= - \infty$, which can never discharge and only 
absorb force from the boundary sites.  In the case of open boundary 
conditions, the sites at the boundary have the same $\alpha$ parameter 
as all other sites in the bulk ($\alpha _{bc}=\alpha$).  In the case of 
free boundary conditions, the sites at the boundary have their $\alpha$ 
parameters modified in order to have the same level of dissipation per
reset event as 
for sites in the bulk.  This latter condition implies 
$\alpha _{bc}=\alpha / (1-\alpha)$, except at corner sites where
$\alpha _{bc,c}=\alpha / (1-2\alpha)$. 
The model with periodic boundary conditions is not critical
\cite{socolar,grass2,middleton},  and will not 
be discussed.
It is probably worthwhile to underline at this point that the model is 
completely deterministic, the only possible source of randomness coming 
from the initial conditions.

After a transient period of  many earthquakes, the model settles into a 
statistically stationary state.  One way to characterize this state is 
to measure statistical properties of the earthquakes.  The size of an 
earthquake, $s$, is defined as the number of resets of the local force
$F_i \rightarrow 0$ in the system in between applications of
the uniform force.  One can also measure the temporal
duration, $t$, in terms of the parallel update, or the radius of
gyration $r$ of the sites which participated in the earthquake event.

\subsection{Finite Size Scaling}
We focus on the probability distribution of earthquake sizes, $s$, in
a system of size $L$, $P_L(s)$.  If the model is critical then this
distribution will have no scale other than the physical extent $L$ and
the lattice constant, which is set to unity.  One ansatz that can
describe critical behavior is the FSS ansatz that was previously used
by OFC:
\begin{equation}
P_L(s) \sim L^{-\beta} G(\frac{s}{L^{D}})
\end{equation}
where $G$ is a suitable scaling function, and $\beta$ and $D$ are
critical exponents describing the scaling of the distribution
function.  As shown in Fig.~1, the model does not exhibit FSS.  In
this figure we chose $D=2$ as the largest possible allowed value.
Still the cutoff in the ``collapsed'' probability distribution moves
to the right as $L$ increases.  Nevertheless, for earthquake sizes
smaller than the cutoff, this figure shows
what appears to be a convergence to a
well-defined power law, $P_L(s) \sim s^{-\tau}$, as $L$ increases with
the power law exponent $\tau =\beta/D \simeq 1.8$ for both
$\alpha=0.18$ and $\alpha=0.21$, and possibly also for $\alpha=0.15$.
\begin{center}
\begin{figure}[hb]
%\protect\vspace{0.3cm}
\narrowtext
 \epsfxsize=3.65truein
\epsffile{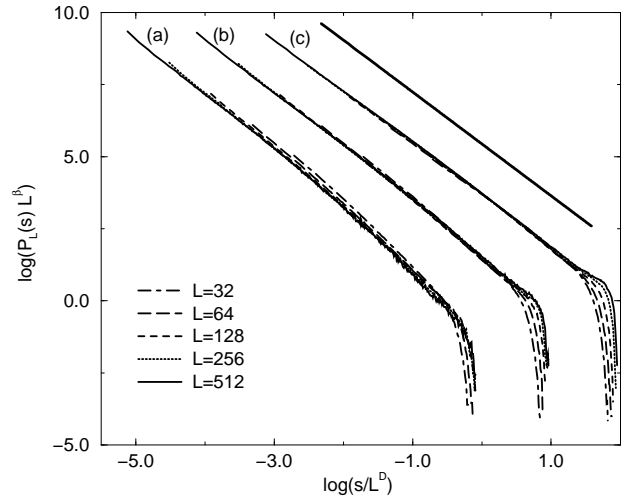}
%\protect\vspace{0.3cm}
\caption[1]{\label{fs_scal}Finite-size scaling plots of $P_L(s)$ in
 systems with open boundary conditions for (a) $\alpha=0.15$, (b)
 $\alpha =0.18$, and (c) $\alpha=0.21$. The critical exponents are
 $D=2$ and $\beta=3.6$, the slope of the straight line is $\tau =
 1.8$. Statistics are derived from at least $10^9 $ avalanches per
 data set. For visual clarity, curves (b) and (c) have been shifted
 along the horizontal axis, $x \rightarrow x+1$ and $x \rightarrow
 x+2$ respectively. }
\end{figure}
\end{center}

\section{Multiscaling Analysis}
The FSS ansatz is only one possible description of critical behavior.
As pointed out some time ago by Kadanoff and co-workers, some SOC
phenomena are better described by a multifractal ansatz, rather than
FSS \cite{kadanoff}.  This form has recently been used to clarify the
behavior of the Bak-Tang-Weisenfeld \cite{soc} model by Stella and
co-workers, who have measured different moments associated with the
distribution \cite{stella}.  For the OFC model, it appears to us that
a clearer picture can be obtained by simply examining the probability
distribution directly.

The multiscaling ansatz postulates for the probability distribution 
function $P_L(s)$ a form 
\begin{equation}
{\log P_L(s/s_o) \over \log (L/l_o)} = F\Bigl( {\log s/s_o \over \log
L/l_o}\Bigr) \quad ,
\end{equation}
where $s_o$ and $l_o$ are parameters which typically (but not always)
reflect phenomena at small scales associated with the lattice \cite{note}.
Usually, a multiscaling analysis consists of choosing these two
parameters in order to get the best collapse for different system
sizes.  This is quite different than FSS where the critical exponents
themselves, reflecting behavior at large scales, must be chosen in an
{\it ad hoc} manner in order to obtain the ``best'' collapse.  We do not
attempt to collapse the data using the multiscaling form of Eq.~2.
Instead, we define $l_o=s_o=1$ to represent the smallest earthquake
which occurs at only one site and involves only one
discharge. Moreover, we define the dimension of an earthquake of size
$s$ in a systems of size $L$ as
$$D_{av}= \log s/\log L \quad ,$$ and we denote with $D_{max}$ the
largest value of $D_{av}$.

In Fig.~2 we show the multiscaled probability distribution according
to Eq.~2, for different system sizes for $\alpha=0.21$.  One observes
immediately that all avalanches have dimension $D_{av}<2$, as required,
with the 
largest dimension $D_{max}$ approaching two as the system size increases.  
Also it is clear that the  shape of the cutoff function is sharpening as the
system size increases.  Since the largest dimension $D_{max}$ cannot be 
larger than two, we can infer from this that the cutoff region narrows 
to the region $D_{av} \rightarrow 2$ and becomes increasingly sharp.
\begin{figure}[hb]
\narrowtext
\protect\vspace{0.3cm}
 \epsfxsize=3.75in
\centerline{\epsffile{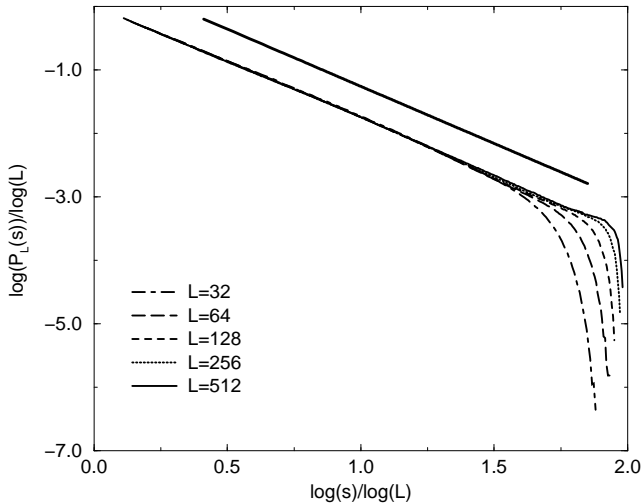}}
\protect\vspace{0.3cm}
\caption[2]{\label{ms_021} Multiscaling plot of $P_L(s)$ for $\alpha=0.21$ 
 and open boundary conditions. The slope of the straight line is 
 $\tau = 1.8$.}
\end{figure}
Note that the increase of $D_{max}$ as $L$ increases is totally
inconsistent with the notion that the OFC model is slightly off
criticality, because in that case one would expect that the relative
size of the largest dissipating
events with respect to the maximum total force allowed in the system
would decrease in larger
systems.  In fact what
happens is exactly the
opposite.  In larger systems a larger fraction of the
total energy can be dissipated in the largest event that occurs,
and $D_{max}$ increases with $L$.  This result
is completely consistent with the nonconservative model being
critical, rather than slightly off criticality.

In order to get more explicit visual information on the probability
distribution we try to subtract out the leading asymptotic term, which we
propose is a power law as suggested by Fig.~2.  This is
\begin{equation}
{\log P_L(s) \over \log (L)}= F(D_{av})= -(\tau D_{av})
+ F_{cutoff}(D_{av}) \quad ,
\end{equation}
where $F_{cutoff}$ in the limit $L \rightarrow \infty$ should be 
constant up to a cutoff near $D_{max}$. 
\begin{center} 
\begin{figure}[hb]
\narrowtext
\protect\vspace{0.3cm}
 \epsfxsize=4.75in
\centerline{\epsffile{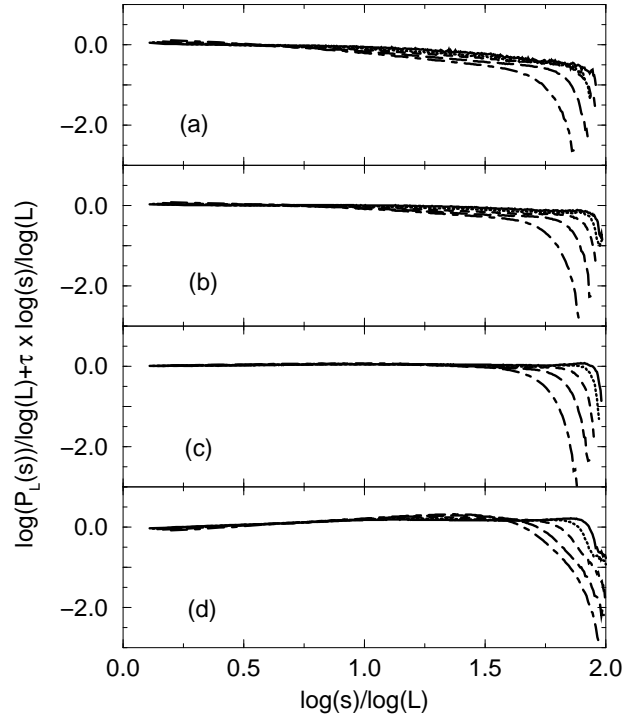}}
\protect\vspace{0.3cm}
\caption[3]{\label{mf_scal}  Plots of $F_{cutoff}$ for  
(a) $\alpha=0.15$, (b) $\alpha =0.18$, (c) $\alpha=0.21$, and 
(d) $\alpha=0.23$; we set $\tau =1.8$. Boundary conditions are open. 
Different curves correspond, from left to right, to $L=32,64,128,256,512$.}
\end{figure}
\end{center}
We get a consistent picture for a range of $\alpha$ values by choosing
$\tau=1.8$, as shown in Fig.~3.  Although it appears for small system
sizes that smaller values of $\alpha$ give a steeper distribution with
a larger value of $\tau$ (so the lines tend to decrease from left to
right rather than remaining horizontal), it is clear from this figure
that as the system size increases all the different values of $\alpha$
appear to reach the same value of $\tau=1.8$, corresponding to a
completely horizontal line in this figure.  The deviation for small
systems is more pronounced for $\alpha=0.15$, and less again for
$\alpha=0.18$, becoming the smallest for $\alpha=0.21$. For $\alpha$
closer to the conservative limit, as shown for $\alpha =0.23$, the
approach to the asymptotic horizontal behavior changes direction.
Namely smaller systems appear to have a smaller exponent $\tau$ than
larger systems, at least for small avalanches.  Thus rather than
having the slope increase as $L$ increases, for $\alpha > \sim .21$,
the apparent slope decreases as $L$ increases, as clearly demonstrated
in this figure.

We ascribe the change of direction to a crossover effect of the conservative
 fixed point.  For $\alpha$ close to 1/4, smaller avalanches behave as
 avalanches would in the conservative system.  It is only the larger
 avalanches that are affected by nonconservative dissipation.  This is
 associated with the fact that as $\alpha$ approaches 1/4, each site
 can topple more and more times in a single earthquake.  For any
 finite value of dissipation ($1-4\alpha$), the maximum number of
 times that a site can reset in an earthquake
 is determined by this dissipation and is
 finite in the limit of large $L$.  However for small systems, the largest
 avalanches are not large enough to be effected by dissipation, and
 the cutoff in the number of times that a site can topple is not
 determined by ($1-4\alpha$) but by $L$.  Then the  cutoff in the
 number of topplings at a given site is the same as in a conservative
 system of the same size.

The results we have described thus far are for the model with open boundary
conditions.  Fig.~4 shows that  the same
behavior occurs for the OFC model with free boundary conditions, with
the same value $\tau=1.8$.  In this case, the asymptotic behavior as $L$
increases is approached for decreasing apparent slope for both
$\alpha=0.17$ and $\alpha=0.20$, as is plainly evident in the figure.
Again the cutoff appears to sharpen as $L$ increases and approach $D_{max}=2$.

\begin{figure}[hb]
\narrowtext
\protect\vspace{0.3cm}
 \epsfxsize=4.5in
\centerline{\epsffile{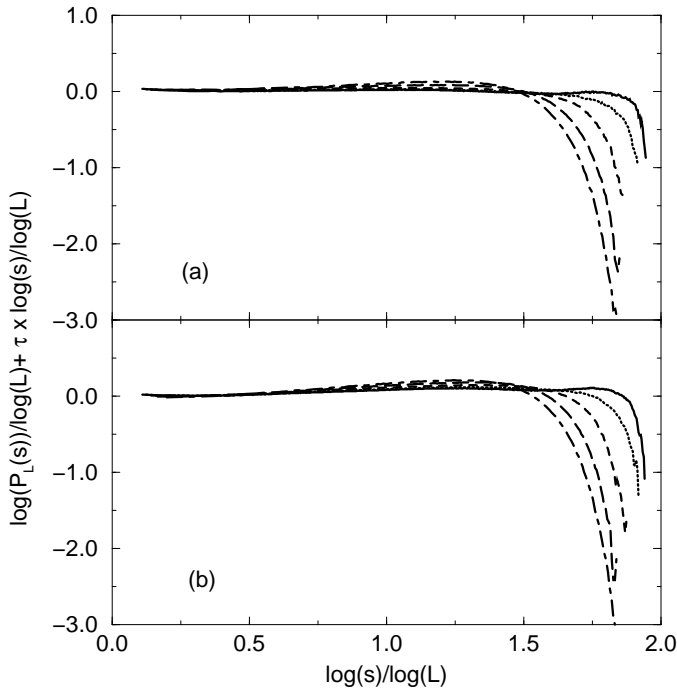}}
\protect\vspace{0.3cm}
\caption[4]{\label{mfs_fbc} Plots of $F_{cutoff}$ in systems with free 
 boundary conditions for (a) $\alpha=0.17$ and (b) $\alpha=0.20$. 
 The exponent has been set to $\tau =1.8$ as before. System sizes are 
 $L=32,64,128,256,512$.}

\end{figure}

Our analysis indicates that the OFC model exhibits SOC with a
universal power law distribution of events $P(s) \sim s^{-1.8}$ for
$\log s < (D_{max} \log L)$.  As shown in Fig.~3, the cutoff gets sharper
and sharper as $L$ increases with $D_{max}$ approaching 2 from below
as $L$ increases for all values of $\alpha$ we have studied.  As
indicated above, the incorrect results obtained by OFC were due to the
fact that there is a strong system size dependence that varies with
$\alpha$ and is not described by FSS.  In addition to a systematic
change in apparent $\tau$ as $L$ increases, the largest dimension of
earthquakes, $D_{max}$, is increasing only slowly towards two, and
probability distribution of avalanche dimensions is getting sharper at
the cutoff.  This means that large avalanches are suppressed in small
systems relative to the total amount of force in the system as compared to a
larger system, and that a FSS ``fit'' for any $L$ range will always
give an apparent $D>2$, which is not allowed.  In this sense the
model appears to violate $FSS$ for all values of $L$.

\section{Conclusions}

The main results of this paper are as follows: The nonconservative
model on a two dimensional lattice self-organizes into a critical
state.  The critical state is robust and universal over a range of
values of the dissipation parameter, $\alpha$, and for different
boundary conditions.  The model does not exhibit finite-size scaling.
The cutoff becomes sharper as $L$ increases with largest earthquakes
of dimension $D_{max}$ approaching two from below.  Nevertheless, the
probability distribution of earthquake sizes is a power law with a
universal exponent $\tau \simeq 1.8$.  This value can be identified
with an exponent $B=\tau -1$ for the distribution of energy dissipated
in earthquakes.  According to the Gutenberg-Richter law this is a
power law with $B=0.8 {\rm \ to \ }1$, completely consistent with our
result.

\medskip

We thank K. Christensen and H. J. Jensen for helpful conversations,
and P. Bak for comments on the manuscript.  S.L. was supported
by the EPSRC through a post-doctoral research fellowship.

\end{multicols}


\begin{references}
\bibitem{note} Throughout this work, in both figures and text,
 $\log$ is the logarithm in base 10.
\end{references}
\end{document}